Near-field Image Transfer by Magneto-Inductive Arrays: a Modal Perspective
by
R.R.A.Syms*, E. Shamonina and L.Solymar

Optical and Semiconductor Devices Group, EEE Dept., Imperial College, Exhibition Road, London SW7 2BT, UK
*Corresponding Author
TEL +44-207-594-6203     FAX +44-207-594-6328     EMAIL r.syms@ic.ac.uk

Abstract

A simple model of near-field pixel-to-pixel image transfer using magneto-inductive arrays is presented. The response of N-dimensional rectangular arrays is first found as an excitation of eigenmodes. This analytical method involves approximating the effect of sources and detectors, and replaces the problem of solving large numbers of simultaneous equations with that of evaluating a sum. Expressions are given for the modal expansion coefficients, and in the low-loss case it is shown that the coefficient values depend only on the difference in reciprocal frequency space of the operating frequency from the resonant frequency of each mode. Analytic expressions are then derived for quasi-optical quantities such as the spatial frequency response, point-spread function and resolving power, and their implications for imaging fidelity and resolution are examined for arrays of different dimension. The results show clearly that there can be no useful image transfer for in-band excitation. Out-of-band excitation allows image transfer. Provided the array is larger than the expected image by at least the size of the point spread function, the effect of the array boundaries may be ignored and imaging is determined purely by the properties of the medium. However, there is a tradeoff between fidelity and throughput, and good imaging performance using thick slabs depends on careful choice of the operating frequency. The approximate analytic method is verified by comparison of exact numerical solution of the full set of coupled equations, and the conditions for its validity are identified.

1.  Introduction

Following seminal work by Veselago [1] and Pendry [2], considerable interest has been shown in the properties of artificial materials that can have negative values of permittivity and permeability by virtue of their physical arrangement as well as their constituents. These 'metamaterials' often involve periodic lattices of resonant elements, which may or may not be coupled to their neighbours. One application is near-field imaging, in the 'perfect lens' geometry proposed by Pendry et al. [3-7]. This arrangement consists of a slab of negative index material, and allows image transfer with a resolution below the diffraction limit by amplification of the evanescent waves that can exist in such materials. Although the idea suffers from some performance limitations [8, 9], experimental confirmation has been provided using a thin layer of silver, which has a negative permittivity at optical frequencies [10, 11]. Since then, analogous devices based on photonic crystals have been proposed [12-14], and operation has been extended to microwave frequencies using arrays of split-ring resonators (SRRs) and wires [15-20].

Near-field imaging of a slightly different kind, involving pixel-to-pixel image transfer or 'canalization', has also been demonstrated at RF frequencies with entirely magnetic metamaterials such as 'Swiss rolls', resonant structures formed from a spiral roll of metal-coated dielectric film [21-23]. Arrays of Swiss rolls can be coupled together magnetically and hence support magneto-inductive (MI) waves [24-27]. Near-field images and the appearance of spatial resonances have both been explained in terms of these waves [28].

Considerable efforts have been made to develop analogous MI lenses based on pairs of stacked planar arrays of SRRs by Freire and Marques [29-32] for applications in magnetic resonance imaging (MRI) [33, 34]. Initially, their operation was explained in terms of the amplification of evanescent fields. However, emphasis has subsequently been placed purely on magneto-inductive effects. It has been convincingly demonstrated that excitation of resonances should be avoided, that imaging may be obtained between the pass-band of the coupled slab system, that the transfer function is not flat and that in-plane coupling reduces fidelity. These conclusions have been verified using detailed calculations involving solution of the full set of coupled equations [35, 36]. Similar predictions and experimental demonstrations have been made using near field imaging systems based on meander resonators [37] and more general transmission-line media [38, 39], with extensions to the optical regime using arrays of metal nanospheres [40].

Pixel-to-pixel image transfer using non-magnetic metamaterials based on arrays of parallel metallic wires has also been extensively investigated [41-43], and extension to optical frequency using periodic metal-dielectric slabs [44] and metallic nanorods has again been proposed [45]. Key advantages of continuous wires over media formed from resonant elements include the very wide potential bandwidth and low loss of the wires, and the ability to form curved image transfer devices very simply [46]. Wire-medium 'endoscopes' are therefore also under investigation as image relays in MRI [47-49].

When considering metamaterial imaging devices, there is often difficulty in reconciling an effective medium approach (in which the properties of the medium are derived from a weighted average of those of the elements) with a full model (involving a microscopic description of each element). The effective medium approach can work extremely well when large numbers of small elements are present, but is difficult to apply to experimental arrangements involving the intermediate numbers of elements that can realistically be manufactured. In this case, boundary effects can be significant and small arrays exhibit complex standing-wave resonances that degrade imaging performance. Furthermore, the number of elements in practical devices is still normally large enough to present difficulties in modeling, since the overall response must be determined by solving many coupled equations. Simulations are carried out on a case-by-case basis, and little exists in the way of performance criteria or design rules. For example, Figure 1 shows several arrangements for near-field imaging, in which signals from a source S are to be transferred to a detector D by a metamaterial array, which might be one- (Fig. 1a), two- (Figs. 1b and 1c) or three-dimensional (Fig. 1d). All have received attention in the previously cited literature. However, it is not clear which dimension of array is best, how large the array should be, how the elements should be arranged, or what the likely performance will be.

The aim of this paper is to provide a simple method of estimating the response of general N-dimensional rectangular lattices used for pixel-to-pixel imaging, and of presenting the result in terms of conventional performance parameters such as the transfer efficiency, spatial frequency response and point spread function. The discussion is focussed on magneto-inductive devices, and coupling to electromagnetic radiation is ignored, but it is hoped that the approach may be applicable to other types of metamaterial and other operating regimes. In Section 2, a general method of estimating the response of a magneto-inductive array is developed, in terms of excitation of a spectrum of eigenmodes. The method replaces the problem of solving large numbers of simultaneous equations with that of evaluating a simple sum, and may therefore be useful in tackling the large-scale problems associated with metamaterial imaging devices. In Section 3, examples are presented for 1D arrays, for which the eigenmodes have simple analytic forms, key optical parameters are deduced and the effect of loss is examined. Comparable results are presented in Section 4 for arrays of higher dimension, and the difference between thin sheets and thicker slabs is highlighted. In Section 5, the approximate analytic method is verified by comparison with numerical solution of the full set of coupled equations, and its regime of validity is discussed. Conclusions are presented in Section 6.

2. <u>Modal excitation theory</u>

In this Section, we present a general theory for the excitation of a magneto-inductive array that allows its response to be found as an expansion of eigenmodes.

*Linear imaging system*

We illustrate the approach using the example of a 1D arrangement of magnetically coupled elements (Figure 1a), now shown in more detail in Figure 2a and in the equivalent circuit of Figure 2b. This arrangement was previously considered in [36]. Here resonant elements containing inductors L (with accompanying resistors R) and capacitors C are coupled to nearest neighbours via mutual inductances M. In pixel-to-pixel imaging, the $n^{th}$ resonant element is also coupled to a source consisting of a voltage source $V_{Sn}$, an inductor $L_{Sn}$ and a resistor $R_{Sn}$ via a mutual inductance $M_{Sn}$ and then to a detector consisting of an inductor $L_{Dn}$ and a resistor $R_{Dn}$ via a mutual inductance $M_{Dn}$. We make no assumptions as to the signs or magnitudes of the mutual inductances, and require only that the elements in the MI array are identical. The sources and detectors may, however, vary, or be resonant. More complex arrangements (in which neighbouring sources or detectors are coupled, or non-nearest neighbour coupling is allowed) are clearly possible, and the array may be finite. However, we neglect these complications, since they do not affect the following argument.

*Governing equations*

Assuming that the current amplitudes in the $n^{th}$ source, element and detector are $I_{Sn}$, $I_n$ and $I_{Dn}$, respectively, the governing equations away from the ends of the array at angular frequency $\omega$ can be found from Kirchhoff's law as:

$$Z_{Sn}I_{Sn} + j\omega M_{Sn}I_n = V_{Sn}$$
$$(j\omega L + 1/j\omega C + R) I_n + j\omega M (I_{n-1} + I_{n+1}) + j\omega M_{Sn}I_{Sn} + j\omega M_{Dn}I_{Dn} = 0$$
$$Z_{Dn}I_{Dn} + j\omega M_{Dn}I_n = 0$$

(1)

Here $Z_{Sn} = j\omega L_{Sn} + R_{Sn}$ and $Z_{Dn} = j\omega L_{Dn} + R_{Dn}$ are the impedances of the $n^{th}$ source and detector. If the sources and detectors are resonant, the reactive contributions to $Z_{Sn}$ and $Z_{Dn}$ can be cancelled at specified frequencies, increasing the relevant currents.

These equations can be collected together and written in matrix form, as $\underline{V} = \underline{Z}\underline{I}$, where $\underline{Z}$ is a matrix of impedances and $\underline{V}$ and $\underline{I}$ are vectors containing voltages and currents (see e.g. [35, 36]). This approach allows a complete solution to be found for the unknown currents by inverting the impedance matrix, as $\underline{I} = \underline{Z}^{-1}\underline{V}$. However, it is intensive in computer time and provides limited physical insight. Here we adopt a perturbation approach common in quantum mechanics that allows analytic results to be deduced very simply. The method involves first reducing the number of equations by absorbing the effect of the sources and detectors into equations describing the resonant elements, and then finding approximate solutions to this reduced set in terms of the array eigenmodes.

We start by re-arranging the upper and lower equations in (1) to give:

$$I_{Sn} = (V_{Sn} - j\omega M_{Sn}I_n) / Z_{Sn}$$
$$I_{Dn} = -j\omega M_{Dn}I_n / Z_{Dn}$$

(2)

Substituting into the central equation in (1) we obtain:

$$(j\omega L + 1/j\omega C + R + \Delta Z_n) I_n + j\omega M (I_{n-1} + I_{n+1}) = U_{Sn}$$

(3)

Here $\Delta Z_n = \omega^2 M_{Sn}^2/Z_{Sn} + \omega^2 M_{Dn}^2/Z_{Dn}$ is an impedance perturbation arising from coupling between the $n^{th}$ array element and the $n^{th}$ source and detector, and $U_{Sn} = -j\omega M_{Sn}V_{Sn}/Z_{Sn}$ is a voltage. The results suggest that the main role of the sources is to impose voltages in the resonant loops, while that of the detectors is to sample the resulting currents. However, both sources and detectors alter the impedances of the array.

*Approximate equations*

We now consider the case when the impedance perturbations are small, a regime previously highlighted in [31] as being necessary for high-quality imaging. In this case $\Delta Z_n$ may be neglected, and Equation 3 solved directly for the array response. The lower of Equations 2 may then be solved for the detector currents. However, the last step is trivial, since these currents are simply proportional to those in the nearby elements. We therefore focus on the first step. To proceed, we write the approximate version of Equation 3 as:

$$(1 - \omega_0^2/\omega^2 - j/Q) I_n + (\kappa/2)(I_{n-1} + I_{n+1}) = I_{Sn}$$

(4)

Here $\omega_0 = 1/(LC)^{1/2}$ is the angular resonant frequency of the elements, $Q = \omega L/R$ is their quality factor (assumed to be high), $\kappa = 2M/L$ is the coupling coefficient and $I_{Sn} = U_{Sn}/j\omega L$ is an effective source current. Equations 4 now represent a set of simultaneous equations, one for each resonant element, that may be written in matrix form as:

$$\underline{M}\,\underline{I} = \underline{I}_S$$

(5)

Here M is a symmetric matrix with diagonal elements $1 - w - j/Q$, where $w = \omega_0^2/\omega^2$ is the square of the normalised reciprocal frequency, and off-diagonal elements $\kappa/2$, and $\underline{I}$ and $\underline{I}_S$ are vectors containing the currents $I_n$ and $I_{Sn}$. Equation 5 has the obvious solution $\underline{I} = \underline{M}^{-1}\underline{I}_S$. Here, however, we re-write it slightly differently, as:

$$(\underline{K} - w - j/Q)\,\underline{I} = \underline{I}_S$$

(6)

Here $\underline{K}$ is a symmetric matrix with unit diagonal elements and off-diagonal coupling terms, and effectively describes the loss-less, unexcited system.

*Loss-less eigenmodes*

We first assume there is no loss, and no excitation. In this case, Equation 6 reduces to:

$$(\underline{K} - w) \underline{I} = \underline{0} \tag{7}$$

The solution of Equation 7 is a set of eigenvectors $\underline{j}_v$ with eigenvalues $w_v$. If these are collected together into matrices $\underline{J}$ (containing the eigenvectors arranged in columns) and $\underline{W}$ (containing the corresponding eigenvalues down the diagonal), Equation 7 can be rewritten as $\underline{K}\,\underline{J} - \underline{J}\,\underline{W} = 0$. This result allows a dyadic spectral expansion of $\underline{K}$, as $\underline{K} = \underline{J}\,\underline{W}\,\underline{J}^{-1}$.

*Expansion into eigenmodes*

We now allow loss and excitation. In this case, a solution for the unknown currents can be attempted as a sum of the eigenvectors of the loss-less system, i.e. as $\underline{I} = \sum_v a_v \underline{j}_v$, where $a_v$ is the coefficient of the mode with eigenvector $\underline{j}_v$. In matrix form this solution may be written:

$$\underline{I} = \underline{J}\,\underline{A} \tag{8}$$

Here $\underline{A}$ is a diagonal matrix of expansion coefficients. Substituting into Equation 6, and using the results above, we get:

$$\{\underline{J}\,\underline{W}\,\underline{J}^{-1} - (w + j/Q)\}\,\underline{J}\,\underline{A} = \underline{I}_S \tag{9}$$

Pre-multiplying by $\underline{J}^{-1}$ we obtain:

$$\{\underline{W} - (w + j/Q)\}\,\underline{A} = \underline{J}^{-1}\,\underline{I}_S \tag{10}$$

We now define a new matrix $\underline{N}$ as $\underline{N} = \underline{W} - (w + j/Q)\,\underline{i}$, where $\underline{i}$ is the identity matrix. The solution for the expansion coefficients is clearly $\underline{A} = \underline{N}^{-1}\,\underline{J}^{-1}\,\underline{I}_S$. However, since $\underline{K}$ is symmetric and real, the eigenvectors $\underline{j}_v$ must form an orthogonal set. If they are also normalised, $\underline{J}^{-1} = \underline{J}^T$. Since $\underline{N}$ is diagonal, $\underline{N}^{-1}$ is easy to evaluate; it is simply a diagonal matrix, whose elements are the reciprocal of the elements of $\underline{N}$. Thus, the expansion coefficient $a_v$ can be written down straight away as:

$$a_v = <\underline{I}_S, \underline{j}_v> / (w_v - w - j/Q) \tag{11}$$

Here $<\underline{I}_S, \underline{j}_v> = \underline{I}_S \cdot \underline{j}_v^*$ is the inner product of $\underline{I}_S$ and $\underline{j}_v$, and is a measure of correlation between the input distribution $\underline{I}_S$ and the mode $\underline{j}_v$. Equation 11 implies that modes will be strongly excited near their resonant frequency, given a suitable excitation pattern.

Equation 11 may be used to find all the mode amplitudes at any frequency. The resulting mode patterns may then be summed to find the overall response. Reverting to angular frequencies, we get:

$$\underline{I} = {}_\nu\Sigma <\underline{I}_S , \underline{j}_\nu> \underline{j}_\nu / \{(\omega_0^2/\omega_\nu^2 - \omega_0^2/\omega^2) - j/Q\}$$

(12)

Equation 12 may be evaluated as a simple sum once the eigenvalues and eigenvectors are known, and can provide the overall response obtained with an arbitrary input. Its form implies that the response will be dominated by those modes for which the denominator is small and increasingly dictated by the operating frequency as the Q-factor rises.

*Generalization to N-dimensions*

Matrix equations may still be constructed even when the array is finite (with $\underline{M}$ being N x N for an N-element array), when non-nearest neighbour couplings are included (by adding further off-diagonal terms), for 2-D and 3D arrangements with arbitrary boundary shapes, and even for aperiodic arrangements. In each case, $\underline{M}$ will be symmetric, and $\underline{K}$ both symmetric and real. Consequently the response of a more complicated array may always be written as an expansion of eigenmodes, and Equation 12 is a general solution. The only difficulty in a more general case is to identify the modes. However, for 1D, 2D and 3D rectangular arrays with nearest neighbour coupling and rectangular boundaries, these have simple analytic forms, and the properties of such arrays may be deduced as a generalization of the results above. The wave-like nature of the eigenfunctions then provides a very simple route to determination of the spatial frequency response.

### 3. Imaging using 1D magneto-inductive arrays

In this Section, we consider the implications of previous results for imaging. We first establish criteria for fidelity, and then consider the behaviour of 1D arrays.

*Imaging fidelity*

Given that the sources and detectors excite and sample currents in those elements nearest to them, a perfect image will be obtained if the source pattern can be transferred through the array without degradation. We must therefore define what is meant by perfect fidelity. This is easiest to do for the linear array of Figure 2. In this case, perfect imaging will be obtained if an excitation pattern $\underline{I}_S$ can simply be transferred onto the array, where it can be sampled by the detectors. Clearly, such a pattern can be expanded as a sum of the array eigenmodes, as $\underline{I}_S = \sum_\nu a_\nu \underline{j}_\nu$. Exploiting orthogonality again, the mode amplitudes $a_\nu$ may be extracted as $a_\nu = <\underline{I}_S, \underline{j}_\nu>$. The 'best' amplitudes may therefore be specified exactly. Unfortunately, the result does not match Equation 11. Comparison shows that amplitudes are scaled during the real excitation process by a factor $S_\nu$, given by:

$$S_\nu = 1 / \{\omega_0^2/\omega_\nu^2 - \omega_0^2/\omega^2 - j/Q\}$$

(13)

Since $S_\nu$ is not constant, perfect imaging is never possible. However, we should still be able to understand the conditions for reasonable performance.

*In-band excitation*

For the linear array, the only free variables are the coupling coefficients and the normalized operating frequency. We first consider the effect of frequency, starting with in-band excitation. If the array is excited at a frequency corresponding to one of the eigenmodes (so that $\omega = \omega_\mu$, say) and the Q-factor is sufficiently high, Equation 11 may be approximated as:

$$a_\mu = jQ <\underline{I}_S, \underline{j}_\mu>$$
$$a_\nu \approx <\underline{I}_S, \underline{j}_\nu> / (\omega_0^2/\omega_\nu^2 - \omega_0^2/\omega_\mu^2) \quad \nu \neq \mu$$

(14)

This result implies that only the mode corresponding to the excitation frequency will be excited significantly, with an amplitude that depends linearly on Q. Its amplitude is also determined by the inner product $<\underline{I}_S, \underline{j}_\mu>$, which expresses its similarity with the excitation pattern. All the other modes will be excited to a certain extent, but in quadrature with the resonant mode. The modes whose eigenfrequencies are closest to $\omega_\mu$ will have the largest amplitude. However, the amplitudes of non-resonant modes will be comparatively small, and as Q rises they can increasingly be ignored. This conclusion has implications for all similar imaging devices, since it suggests that operation within the frequency band supporting propagating waves will tend to result mainly in the excitation of resonances. These findings are entirely in agreement with the literature.

*Out-of-band excitation*

Having deduced that fidelity in imaging cannot easily be combined with resonant gain, we are left with the possibility of out-of-band excitation. In this case, if the Q-factor is large enough, Equation 11 may be approximated for all modes simply as:

$$a_\nu \approx <\underline{I}_S, \underline{j}_\nu> / (\omega_0^2/\omega_\nu^2 - \omega_0^2/\omega^2)$$

(15)

Now, the coefficient values depend only on the difference in reciprocal frequency space of the operating frequency from the resonant frequency of each mode.

Similarly, $S_\nu$ may be written as:

$$S_\nu = 1 / (\omega_0^2/\omega_\nu^2 - \omega_0^2/\omega^2)$$

(16)

Although $S_\nu$ is presented here as a function of angular frequency, spatial and temporal frequencies are related by the dispersion equation. As we will show, $S_\nu$ must therefore represent the spatial frequency response (SFR).

*Spatial frequency response*

The SFR may be found for a linear array as follows. If the array is infinite, its eigenmodes are the continuous set of travelling current waves $I_n = I_0 \exp(\pm jnka)$, where k is the propagation constant at angular frequency $\omega$, n is an integer and a is the lattice period. In this case, the loss-less dispersion relation is [24]:

$$\omega_0^2/\omega_\nu^2 = 1 + \kappa \cos(ka)$$

(17)

Figure 3a shows a typical dispersion diagram, obtained from Equation 17 by assuming a negative coupling coefficient $\kappa = -0.25$, which requires the elements to be arranged in the planar configuration. Propagation is band-limited, and obtained only over the frequency range between $\omega_0^2/\omega_\nu^2 = 1 + \kappa$ and $\omega_0^2/\omega_\nu^2 = 1 - \kappa$. The curve is slowly varying, and flattest near the band edges.

In this case, S is a continuous function, found by combining Equations 16 and 17 to get:

$$S(k) = 1 / \{1 + \kappa \cos(ka) - \omega_0^2/\omega^2\}$$

(18)

If, on the other hand, the array is finite, the eigenmodes are standing waves. For an N-element line, the allowed modes must satisfy the resonance condition $ka = \nu\pi/(N + 1)$, where $\nu$ is an integer with allowed values 1, 2 …N. The allowed values of ka are then discrete points on Figure 3a, and Equation 18 must be replaced with the discrete function:

$$S_\nu = 1 / \{1 + \kappa \cos[\nu\pi/(N+1)] - \omega_0^2/\omega^2\}$$

(19)

Equations 18 and 19 are clearly analogous. The availability of analytic forms for the spatial frequency response now allows a conscious choice of the design and operating parameters. For a 1-D array, these are simply $\kappa$ and $\omega/\omega_0$, respectively.

As we have shown, any useful operation must be out of band, for example, at frequencies defined by the thick or thin straight lines in Figure 3a. In the thick-line case, the denominator in Equation 18 is small only when $ka/\pi$ is close to unity, i.e. for high spatial frequencies. Consequently, we would expect low spatial frequencies to be suppressed. Similarly, for the full lines it is small only at low spatial frequencies. Thus, we cannot simultaneously transfer both low and high spatial frequencies, and must choose one or other. For conventional imaging, good transmission of low spatial frequencies is required. We therefore now focus on the full-line cases.

Assuming that the operating frequency lies above the upper band edge, we may define the angular frequency used as $\omega_0^2/\omega^2 = 1 + \kappa - \delta$, where $\delta$ is a deviation in reciprocal frequency space. Each of the thin lines corresponds to operation at a particular value of $\delta$, which increases as the line moves away from the upper band edge. Using this definition of the operating point, the spatial frequency response can be written:

$$S(k) = 1 / \{\delta + \kappa [\cos(ka) - 1]\}$$
(20)

Figure 3b shows the spatial frequency responses obtained from (20), for the same values of $\kappa$ and $\delta$ as before. In each case, the response is low-pass. As $\delta$ increases, the peak in response reduces, but the spatial frequency bandwidth increases, so there is a trade-off between image brightness and fidelity of reproduction. The bandwidth also increases if $|\kappa|$ is reduced, implying that lateral coupling is inherently deleterious to image quality, in agreement with earlier conclusions (see e.g. [32]). Similar spatial responses (with different brightness) can be obtained for different values of $\kappa$, provided $\delta$ is scaled appropriately.

Using a small angle approximation for the cosine, Equation 20 reduces to:

$$S(k) = 1 / \{\delta - \kappa (ka)^2/2\}$$
(21)

Numerical evaluation shows that Equation 21 is a good approximation to Equation 20, and may be used to obtain simple analytic estimates of performance.

*Point spread function*

In the spatial domain, the response of an imaging system is described by the point-spread function (PSF). For a 1-dimensional array, the PSF can be found as the response to unit excitation of a single element, such as element zero. Using the symbols $P_n$ instead of $I_n$ for the currents (to denote the PSF) we must solve the equations:

$$(1 - \omega_0^2/\omega^2) P_0 + (\kappa/2)(P_{-1} + P_{+1}) = 1$$

$$(1 - \omega_0^2/\omega^2) P_n + (\kappa/2)(P_{n-1} + P_{n+1}) = 0 \text{ for } n \neq 0$$
(22)

The general solutions to the lower equation are travelling waves, $P_n = P_0 \exp(-jnka)$ for n > 0 and $P_n = P_0 \exp(+jkna)$ for n < 0. Substituting into the upper equation and making use of the dispersion equation we obtain $P_0 = j/\{\kappa \sin(ka)\}$, so that the PSF must be:

$$P_n = j \exp(-j|n|ka) / \{\kappa \sin(ka)\}$$
(23)

This expression does not at first sight resemble a conventional point-spread function, since its modulus appears to be constant. However, we note that ka is purely imaginary outside the band, i.e. at the temporal frequencies at which useful image transfer can occur. In this case, we can write $ka = -jk''a$, and the PSF becomes:

$$P_n = -\exp(-|n|k''a) / \{\kappa \sinh(k''a)\}$$
(24)

The PSF is therefore bounded as expected, and decays exponentially on either side of the excitation. The decay rate is determined by the value of k''a, which will become larger at frequencies further from the operating band. Equation 24 remains a good approximation for centrally excited arrays of finite size, provided the decay rate is such that the current amplitude is small at the array edges. Under these circumstances, the effect of the array boundaries may be ignored and imaging performance is determined entirely by the properties of the medium. To achieve this result, the image should be placed centrally, and the array should be larger than the expected image size by at least the width of the PSF.

For compatibility with previous results we now express the PSF in terms of the deviation $\delta$ from the band edge. Simple manipulation yields $\cosh(k''a) = (1 - \delta/\kappa)$, allowing k''a and sinh(k''a) to be found. Clearly, when $\delta$ is zero, the decay rate of the exponential is zero. The PSF is then entirely flat and there can be no image transfer at all. As $\delta$ increases, the decay rate also increases. A point-like image can now be transferred to the array and this image becomes increasingly sharp. However, its peak amplitude reduces, highlighting again the tradeoff between image fidelity and brightness. This behaviour is illustrated in Figure 3c, which shows the PSF plotted on a logarithmic scale, for the same values of $\kappa$ and $\delta$ as before. These results are again in excellent agreement with numerical calculations in [36].

*Transform relation*

In conventional optics, the SFR of an imaging system is related to the PSF by a transform. We therefore now show that the linear magnetoinductive array obeys similar rules, by expanding the point spread function as a spectrum of eigenmodes. The algebra is simplest if the array is first considered to be finite, and its size is then allowed to tend to infinity. We therefore assume that the index n ranges from -A to +A, so the total number of elements is N = 2A+1. In this case, the normalised eigenmodes are the cosines $\sqrt{\{2/(N + }$ 1)} $\cos(nk_\nu a)$, where $k_\nu a$ has the discrete values $\nu\pi/(N+1)$ as before. Using this spectrum of modes, the spatial frequency response can be written as the discrete cosine transform:

$$S_\nu = {}_{n=-A}\Sigma^A P_n \cos(nk_\nu a)$$

(25)

Substituting for the PSF, we get:

$$S_\nu = \alpha \; {}_{n=-A}\Sigma^A \exp(-|n|k''a) \cos(nk_\nu a)$$

(26)

Here $\alpha = -1 /\{\kappa \sinh(k''a)\}$. Allowing A to tend to infinity, the summation may be evaluated after some manipulation as $S(k) = -1 / \{\kappa [\cosh(k''a) - \cos(ka)]\}$. Finally, using the dispersion equation we obtain:

$$S(k) = 1 / \{1 + \kappa \cos(ka) - \omega_0^2/\omega^2\}$$

(27)

Since Equation 27 is in full agreement with earlier results, it does indeed appear that the magneto-inductive array obeys the rules of conventional imaging.

*Loss*

We now consider briefly the effect of loss in the resonant elements. If the Q-factor is now finite, Equation 20 modifies to:

$$S(k) = 1 / \{\delta + \kappa [\cos(ka) - 1] - j/Q\}$$

(28)

Neglect of the final term will cause the largest inaccuracy at zero spatial frequency, when $ka = 0$. In this case, $S(k) = 1 / \{\delta - j/Q\}$. The earlier loss-less formulae will therefore represent a good approximation if $\delta > 1/Q$. Since typical experimental Q-factors are of order 100, significant effects are likely to be seen only for small $\delta$. We illustrate this point in Figure 4a, which shows the modulus of the SFR for $\kappa = -0.25$ and $\delta = 0.01$, for lossy 1-D arrays with different values of Q. For $Q > 100$, there is little difference in the response. Figure 4b shows the moduli of the corresponding point spread functions, which vary exponentially for all Q and simply reduce in peak amplitude and narrow as Q falls. These results imply that imaging quality actually rises as losses increase, due to a reduction in lateral propagation distance. However, this improvement is counterbalanced by a reduction in image brightness.

*Resolution*

The resolution of the array can be defined in terms of its ability to form a separated image of two point objects of equal amplitude. As we have seen, the point spread function may be written as $P(n) = \alpha \exp(-|n|k''a)$. A single point object located at $n = +n_O/2$ will therefore produce a response $I_+(n) = \alpha \exp(-|n - n_O/2|k''a)$, while a similar object at $n = -n_O/2$ will yield $I_-(n) = \alpha \exp(-|n + n_O/2|k''a)$. Their combined image can be found by

superposition, as $I(n) = I_-(n) + I_+(n)$. Figure 5a shows the image obtained using a loss-less array, assuming $n_O = 4$ and $\kappa = -0.25$, for different values of $\delta$. As $\delta$ rises, the overall amplitude falls, but the relative depth of the valley at $n = 0$ between the separate point images clearly increases.

Adopting a Rayleigh-like criterion, we might define the two points as being resolved if the height of the valley is less than a given fraction $\eta$ of the peak, or if:

$$2 \exp(-n_O k''a/2) < \eta \tag{29}$$

The minimum separation between resolvable points is then:

$$n_O > (2/k''a) \log_e(2/\eta) \tag{30}$$

Using previous results, $n_O$ may be obtained in terms of the reciprocal frequency deviation term $\delta$. Figure 5b shows the variation of the minimum resolvable object separation $n_O$ with $\delta$, again assuming $\kappa = -0.25$ and different values of the separability criterion $\eta$. Pixel-scale resolution is obtained when $\delta$ is approximately unity.

4. <u>Imaging using magneto-inductive arrays of higher dimension</u>

In this Section, we extend the analysis to 2D and 3D arrays, beginning with thin sheets (Figure 1c) and thin slabs (Figure 1b) and finally thick slabs (Figure 1d). As we shall see, thin sheets and slabs behave very differently, while thick slabs combine the major characteristics of each type.

*Imaging using 2D sheets*

Figure 6 shows an imaging arrangement in which a two-dimensional sheet of magneto-inductive material containing a set of identical resonant elements is interposed between a set of sources and detectors, which are coupled to their nearest neighbours in the array as before. To obtain a symmetric response, elements in the array must be coupled by equal mutual inductances M in the x- and y-directions and separated by equal distance a. In this case M and the corresponding coupling coefficient $\kappa$ can only be negative.

The relevant equations are entirely analogous to those of the 1D case. In an infinite array, the eigenmodes are the two- dimensional current waves $I_{n, m} = I_0 \exp(-jnk_xa) \exp(-jmk_ya)$, where $k_x$ and $k_y$ are propagation constants in the x- and y-directions, n and m are integers denoting the element position and a is the lattice period. In this case, the dispersion equation is [25]:

$$\omega_0^2/\omega_v^2 = 1 + \kappa \{\cos(k_xa) + \cos(k_ya)\}$$
(31)

This result implies that the dispersion characteristic is now a symmetric curved sheet, whose uppermost point lies at $\omega_0^2/\omega_v^2 = 1 + 2\kappa$. Figure 7a shows a typical characteristic, calculated assuming $\kappa = -0.25$.

By analogy with previous results, the spatial frequency response $S(k_x, k_y)$ may also be written down directly, as:

$$S(k_x, k_y) = 1 / \{1 + \kappa [\cos(k_xa) + \cos(k_ya)] - \omega_0^2/\omega^2\}$$
(32)

Since $k_x$ and $k_y$ are similarly represented, the spatial frequency response must be equal in x- and y-directions. In the array is finite, the allowed values of $k_xa$ and $k_ya$ become a set of discrete points as before. Once again, high-quality imaging may only be obtained if the operating frequency lies just above the upper band edge. Defining the frequency as $\omega_0^2/\omega^2 = 1 + 2\kappa - \delta'$, where $\delta'$ is a modified deviation in reciprocal frequency space, the SFR can be written as:

$$S(k_x, k_y) = 1 / \{\delta' + \kappa [\cos(k_xa) + \cos(k_ya) - 2]\}$$
(33)

Figure 7b shows the spatial frequency response, again calculated assuming $\kappa = -0.25$, and now assuming that the operating frequency is defined as $\delta' = 0.05$. The similarity of this expression to the earlier 1D result implies that similar imaging behaviour must be obtained, but at the slightly different operating point used in two dimensions.

By symmetry, the point-spread function may then be found as before, as the function:

$$P_{n,m} = j \exp(-j[|n| + |m|]ka) / \{2\kappa \sin(ka)\}$$
(34)

Figure 7c shows the point spread function, again calculated assuming $\kappa = -0.25$ and $\delta' = 0.05$. The result is again a symmetric 2D equivalent of the 1D result, suggesting that there is little qualitative difference between the imaging performance of lines and sheets.

*Imaging using thick 2D slabs*

Figure 8 shows a further arrangement in which a two-dimensional slab of magneto-inductive material is interposed between a set of sources and detectors. Arrays of this type have previously been considered by a number of authors. Here, the array is assumed to contain an arbitrary (but finite) number $N_z$ of lines; sources are assumed to be coupled to the first line of elements and the detectors to the last line, and the details of these couplings are as before. Within the array, nearest neighbours are assumed to be separated by distances $a_x$ and $a_z$ and coupled by mutual inductances $M_x$ and $M_z$ as shown. There is some freedom to choose the signs of $M_x$ and $M_z$; here we assume $M_x$ is negative and $M_z$ is positive, since this arrangement may easily be extended to three dimensions.

In an infinite two-dimensional MI array, the loss-less dispersion equation can again be found by assuming wave solutions to the circuit equations, as [25]:

$$\omega_0^2/\omega_v^2 = 1 + \kappa_x \cos(k_x a_x) + \kappa_z \cos(k_z a_z)$$
(35)

Here $\kappa_x = 2M_x/L$ and $\kappa_z = 2M_z/L$ are coupling coefficients in the x- and z-directions, and $k_x$ and $k_z$ are the corresponding propagation constants.

Here, however, the array is finite in the z-direction, and this aspect introduces a major qualitative difference from the 2D sheet. The 2D slab eigenmodes must have the form of standing waves in this direction and hence must be written as $I_{n,o} = I_0 \exp(-jnk_x a_x) \sin(ok_z a_z)$, where n and o are integers denoting the element position, $k_z a_z = \mu\pi/(N_z+1)$ and $\mu$ is an integer with allowed values 1, 2 … $N_z$. In this case, the dispersion equation becomes:

$$\omega_0^2/\omega_v^2 = 1 + \kappa_x \cos(k_x a_x) + \kappa_z \cos[\mu\pi/(N_z+1)]$$
(36)

When plotted as a function of $k_x a_x$, the dispersion relation becomes a set of bands, one for each value of $\mu$. This conclusion is illustrated in Figure 9a, which shows an example characteristic obtained for the parameters $\kappa_x = -0.05$, $\kappa_z = 0.5$ and $N_z = 6$. Here there are clearly six bands. By analogy with previous results, it should be possible to obtain pixel-to-pixel image transfer using an operating frequency near the top of any of them. In [36], this effect was explored numerically for a bi-layer system, which has just two bands. It was shown that image transfer could be achieved at frequencies near the top of either of the bands, provided the coupling coefficient $\kappa_z$ was large enough to open a gap between them.

Even if the bands are separate, they must move closer together as $N_z$ rises. Since the operating frequency must be placed increasingly close to the peak of the desired band to ensure that modes from this band are predominantly excited, the lateral coupling coefficient must be reduced to ensure a reasonably flat transfer function. If this can be done, the simplest procedure is to operate near the top of the upper band, as shown by the horizontal line in Figure 9a. Operating here, the primary effect of excitation should be to generate a spectrum of modes with $\mu = N_z$. In the x-direction, these represent arbitrary travelling waves, but in the z-direction the current variation must be the highest order standing resonance of the slab, and the image at the output should similar to the pattern impressed at the input.

For the uppermost band, the mathematics is much as before. The dispersion equation is:

$$\omega_0^2/\omega_v^2 = 1 + \kappa_x \cos(k_x a_x) + \kappa_z \cos[N_z\pi/(N_z+1)]$$

(37)

Consequently, the spatial frequency response for this band alone must be:

$$S(k_x) = 1 / \{1 + \kappa_x \cos(k_x a_x) + \kappa_z \cos[N_z\pi/(N_z+1)] - \omega_0^2/\omega^2\}$$

(38)

The edge of the this band lies at $\omega_0^2/\omega^2 = 1 + \kappa_x + \kappa_z \cos[N_z\pi/(N_z+1)]$. Defining the operating point as $\omega_0^2/\omega^2 = 1 + \kappa_x + \kappa_z \cos[N_z\pi/(N_z+1)] - \delta''$, where $\delta''$ is a further deviation in reciprocal frequency space, the spatial frequency response becomes:

$$S(k_x) = 1 / \{\delta'' + \kappa_x [\cos(k_x a_x) - 1]\}$$

(39)

This expression is clearly analogous to previous results obtained using single lines, implying that similar results may be obtained using slabs, but at the slightly different frequencies that follow from to the replacement of $\delta$ by $\delta''$.

The analysis above is clearly a simplification, since the effect of any modes that are excited in other bands must also be taken into account. If this is done, both the spatial frequency response and the point spread function must vary with distance through the slab. Given the form of the modal expansion coefficients, the most significant unwanted modes lie the second highest band, and their longitudinal variation must lead to some cancellation of the desired modes at the output. To illustrate this, Figure 9b shows the variation with n of the modulus of the point-spread function P(n, o) at the input and output of the slab, for the parameters $\kappa_x = -0.05$, $\kappa_z = 0.5$, $N_z = 6$ and $\delta'' = 0.001$. At the input (o = 1), the current decreases exponentially on either side of the excitation point. At the output (o = 6) the peak of the PSF has reduced, primarily due to the effect of exciting unwanted modes.

Figure 9c shows the variation with longitudinal position o of the modulus of the peak of the point-spread function. If only modes in the uppermost band were excited, we would expect this variation to follow a sinusoidal standing wave pattern (points labelled 'Highest band'), as the image is transferred through the slab. However, excitation of modes in the second highest band has introduced asymmetry (points labelled 'All bands'), increasing the input amplitude and reducing the output. The only solution is to move the operating point closer to the upper band, so that mode amplitudes are preferentially enhanced in this band. However, this will in turn degrade the spatial frequency response. Effects of this type suggest that any potential benefits from the use of thick slabs (e.g., an increase in image transfer distance) are likely to be counteracted by a reduction in image quality.

*Imaging using 3D slabs*

Figure 10 shows a final arrangement in which a three-dimensional slab of magneto-inductive material is interposed between a set of sources and detectors. From previous results, we would expect their operation to be an amalgam of the behaviour of 2D sheets and slabs. The dispersion characteristic will split into a stacked set of curved surfaces, each similar to the single surface obtained for a 2D sheet. Image transfer may be obtained by operating at a frequency close to the highest point of one such surface, and the point spread function for modes in a single band will then be analogous to that obtained for a 2D sheet. For the upper band alone, the spatial frequency response $S(k_x, k_y)$ is:

$$S(k_x, k_y) = 1 / \{1 + \kappa_x [\cos(k_x a_x) + \cos(k_y a_x)] + \kappa_z \cos[N_z \pi/(N_z+1)] - \omega_0^2/\omega^2\}$$
(40)

The edge of the this band lies at $\omega_0^2/\omega^2 = 1 + 2\kappa_x + \kappa_z \cos[N_z\pi/(N_z+1)]$. Defining the operating point as $\omega_0^2/\omega^2 = 1 + 2\kappa_x + \kappa_z \cos[N_z\pi/(N_z+1)] - \delta'''$, where $\delta'''$ is a further deviation in reciprocal frequency space, the spatial frequency response becomes:

$$S(k_x, k_y) = 1 / \{\delta''' + \kappa_x[\cos(k_x a_x) + \cos(k_y a_x) - 2]\}$$
(41)

This result clearly has a similar form to previous analogous expressions, implying the possibility of similar image transfer. However, as in a 2D slab, performance will be degraded by the excitation of modes in any adjacent band.

Consequently, we would again expect the point spread function to alter through the slab. Figures 11a and 11b shows PSFs obtained at the input and output, respectively, of a thick slab with $\kappa_x = \kappa_y = -0.05$, $\kappa_z = 0.5$ and $N_z = 6$ and $\delta''' = 0.001$. At the input (o = 1), the current decreases exponentially in both directions on either side of the excitation point. At the output (o = 6), the PSF has a qualitatively similar shape. However, the central peak of the transferred pattern has again reduced significantly and the PSF has broadened.

We illustrate more general 3D slab imaging performance with a single example that highlights the effect of slab thickness. Figure 12 shows images obtained at the output of different loss-less 3D slab magneto-inductive arrays of a letter 'M', defined as a centrally-placed line object measuring 16 units by 8 units. Each array has the coupling coefficients $\kappa_x = \kappa_y = -0.05$, and $\kappa_z = 0.5$, and the slab thicknesses are $N_z = 2$ (Figure 12a), $N_z = 4$ (Figure 12b), and $N_z = 6$ (Figure 12c). The operating point is defined by taking $\delta''' = 0.01$. In each case, the image is successfully transferred; however, there is a steady degradation in the image brightness and quality as the slab thickness rises.

*Design rules*

Based on the discussion above, design rules for N-dimensional magneto-inductive near-field imaging devices may be summarised as follows. Symmetric arrays should be used to obtain a symmetric response. Lateral coupling should be minimised to extend the spatial frequency response as far as possible. The longitudinal coupling should be maximised and the slab thickness should be minimised to increase the spacing between bands. The operating frequency should be chosen to lie just above one of the band edges. Clearly, even the simplest equation for the spatial frequency response (e.g. Equation 18) contains the operating frequency, so the near-field MI imaging devices of this type cannot possibly be achromatic. However, modest narrow-band performance appears possible, and might be sufficient for systems such as MRI.

5.  Comparison with exact solution

In this section, we consider the accuracy of the approximate theory in comparison with a full numerical solution obtained by solving the matrix equation $\underline{V} = \underline{Z}\,\underline{I}$ for the array together with sets of sources and detectors whose effects are no longer negligible.

*Numerical model*

A 1D planar array is assumed in the geometry of Figure 2, containing 101 resonant elements numbered -50 to +50 with resonant frequency $\omega_0$ and a quality factor Q. The coupling coefficient κ is assumed to be negative. Excitation is by a single central source at element zero, which is only coupled to one element in the array. The array currents are found using three different approaches, which cause successively increasing loading $\Delta Z_n$ on the array. Method 1 involves simple calculation of the currents, with no detectors present, Method 2 uses sampling by a single detector placed at different locations near element zero, and Method 3 uses sampling by a set of fixed detectors spanning the entire array. In each case, the mutual inductances $M_S$ and $M_D$ are positive, and defined in terms of coupling coefficients $\kappa_S = 2M_S/L$ and $\kappa_D = 2M_D/L$. The sources and detectors are also assumed to be resonant, but at the operating frequency ω, and have Q-factors $Q_S$ and $Q_D$.

Many numerical calculations were performed, with different parameter combinations. The coupling coefficient κ in the array was fixed at -0.25, while the Q-factor of the elements was varied from 10 to 1000. The coupling coefficients $\kappa_S$ and $\kappa_D$ to the sources and detectors were taken as being equal, and varied from 0.01 to 2, while the corresponding quality factors $Q_S$ and $Q_D$ were taken as being equal to Q. The operating frequency was defined in terms of the parameter δ, which was varied from 0.05 to 2.

*Numerical results*

The three approaches were used to find the point spread function under different conditions, allowing the following general conclusions to be reached. Out-of-band, the PSF is always a function that decays exponentially on either side of the excitation point. However, the peak amplitude and the decay rate both depend on the exact arrangement and model parameters. For example, using Method 1 with $\kappa_S = +0.025$ and Q = 100, the results in Figure 3c (which shows the loss-less PSF) were reduced almost exactly for all values of δ, merely provided the currents in the array are corrected by a constant factor. Similar results were obtained with Method 2. However, larger discrepancies were obtained using Method 3, especially for large $\kappa_S$ and Q or small δ, and the decay rate of the PSF was found to increase significantly. For example, Figure 13 shows point spread functions calculated with $\kappa_S = +0.025$ and Q = 1000, for δ = 0.05 (Figure 13a), δ = 0.1 (Figure 13b) and δ = 0.2 (Figure 13c). Here, peak amplitudes have all been normalised to unity for comparison. In each case, the approximate solution is in excellent agreement with the prediction of Methods 1 and 2; however, agreement with Method 3 is worse for small δ.

*Analytic explanation*

These conclusions may be explained using simple analysis, as follows. Including the effect only of loading by a single source at element zero (Method 1), the equations that must be solved to find the loss-less PSF are modified versions of Equations 22:

$$(1 - \omega_0^2/\omega^2 - j\Delta_0) P_0 + (\kappa/2)(P_{-1} + P_{+1}) = 1$$
$$(1 - \omega_0^2/\omega^2) P_n + (\kappa/2)(P_{n-1} + P_{n+1}) = 0 \text{ for } n \neq 0$$

(42)

Here $\Delta_0 = \Delta Z_0/\omega L$ is a normalised impedance perturbation due to the source. If (as here) the source is resonant, $\Delta$ may be written alternatively as $\Delta_0 = \kappa_S^2 Q_S/4$, so the normalised perturbation increases with both $\kappa_S$ and $Q_S$. Equations 42 have an analytic solution comparable to Equation 23, namely:

$$P_n = j \exp(-j|n|ka) / \{\kappa \sin(ka) + \Delta_0\}$$

(43)

The term $ka$ may clearly be replaced with $k''a$ to obtain a modified version of Equation 24. However, the difference between Equations 23 and 43 only lies in the denominator. Consequently, the PSF must decay exponentially on either side of the excitation point, at the same rate as in the unloaded case. Only the amplitude alters, and the relative magnitude of the change depends on the size of $\Delta$ compared with $\kappa \sin(ka)$. The fractional change will be small if $\Delta_0$ is relatively small. For $\kappa_s/\kappa = -0.1$ (as here), this will be the case if $Q_S < 100$ and $\sin(ka)$ is greater than unity. The last condition only requires that $\delta$ is sufficiently large.

When a single detector is used, at the same location as the source, the effect is simply to modify the value of $\Delta_0$. For example, if $\kappa_D = \kappa_S$ and $Q_D = Q_S$, $\Delta_0$ will double. Consequently, we would expect Equation 43 to be valid in this case as well. As a result, the decay rate of the PSF will again be unaltered, and only the peak amplitude will change slightly. If the detector is now moved (Method 2), we would again expect the effect to be small.

When multiple detectors are used with a single source (Method 3), the effect is to modify Equations 42 by inserting additional perturbations into <u>all</u> of the equations as follows:

$$(1 - \omega_0^2/\omega^2 - j\Delta_0) P_0 + (\kappa/2)(P_{-1} + P_{+1}) = 1$$
$$(1 - \omega_0^2/\omega^2 - j\Delta_1) P_n + (\kappa/2)(P_{n-1} + P_{n+1}) = 0 \text{ for } n \neq 0$$

(44)

For example, if $\kappa_D = \kappa_S$ and $Q_D = Q_S$, $\Delta_0 = 2\Delta_1$. Examining Equations 44, we see that the effect of loading by a single source and multiple detectors is to insert additional loss into each element, and that the loss is almost uniform. However, the resulting perturbation depends linearly on $Q_S$ and $Q_D$ (in contrast to the array, where similar perturbations are inversely proportional to Q). Consequently, we would expect any change in the PSF to mimic that previously shown in Figure 4b for the case of a lossy array, and this conclusion is confirmed numerically in Figure 13.

Consequently, the approximate theory should give accurate results for any combination of loss, sources and detectors, under the following conditions. The Q-factors of the elements in the array should be relatively high, the Q-factors of the sources and detectors should be relatively low, the coupling coefficients between the sources and detectors should also be relatively small, and the frequency deviation parameter δ should not be too small.

6.  Conclusions

An approximate but general theory has been presented for excitation of magneto-inductive arrays of various dimensions as an expansion of eigenmodes, avoiding the problem of solving the large number of simultaneous equations associated with such arrays. The method allows a simple estimate of the modal expansion coefficients. Provided the Q-factor is high enough, and individual modes are not resonant, the coefficient values depend only on the separation in reciprocal frequency space of the operating frequency from the resonant frequency of each mode. For rectilinear arrays, the harmonic form of the eigenmodes then allows a simple connection to the spatial frequency response obtained in imaging. The model has been compared with a full numerical solution, and has been shown to give good results provided high-Q arrays are weakly coupled to sources and detectors with moderate Q-factor.

The approach has been used to estimate the performance of magneto-inductive arrays as near-field pixel-to pixel image-transfer devices. In-plane coupling is shown to degrade fidelity. When operated in-band, such coupling can lead to the excitation of resonances, which then dominate the response. Out-of-band, it leads to a degradation of the spatial frequency response and the point-spread function. The best results are obtained if the operating frequency is chosen to lie just above the upper band edge, and if a single eigenmode can be excited in the direction of propagation. In this case, the point spread function is bounded, the effect of the array boundaries vanishes and performance is determined purely by simple properties of the medium and the operating frequency.

Imaging performance is degraded in thick slabs as the bands crowd closer together. The lateral coupling coefficient should therefore be small (so that the spatial frequency response is flat) and the longitudinal coupling coefficient should be large and the slab thickness small (so that the bands are widely separated). In this case, the form of the array is effectively a set of short, isolated magneto-inductive 'wires', not unlike a wire-based imaging medium. Although images may clearly be transferred, the spatial frequency response is strongly dependent on the operating frequency. Consequently, the development of arrangements that allow broadband operation and achromatic performance remains a significant challenge.

Although the method has been used to estimate the response of coupled systems with particular lattice arrangements, boundary shapes and target functions, it is hoped that it may be useful as a rapid way of estimating the response of other metamaterial systems, or their performance in other potential applications.

8. <u>Figures</u>

1. Magneto-inductive near-field pixel-to-pixel imaging systems based on a) 1D, b) and c) 2D and d) 3D rectilinear arrays.
2. a) Arrangement and b) equivalent circuit of a 1D magneto-inductive imaging system.
3. a) Dispersion diagram for a loss-less 1D magneto-inductive array with coupling coefficient $\kappa = -0.25$; b) spatial frequency response, and b) point spread function, for different values of the normalised frequency deviation parameter $\delta$.
4. a) Spatial frequency response and b) point spread function of a lossy 1-D magneto-inductive array, for $\kappa = -0.25$ and $\delta = 0.01$ and different values of Q-factor.
5. a) Images of two point objects located at $n = \pm 2$ obtained using a loss-less 1-D array, calculated assuming $\kappa = -0.25$ and different values of $\delta$; b) variation of minimum resolvable object separation with $\delta$, for different values of the separability criterion $\eta$.
6. Arrangement of a 2D sheet magneto-inductive imaging system.
7. a) Dispersion diagram for a loss-less 2D sheet magneto-inductive array with $\kappa = -0.25$; b) spatial frequency response and c) point spread function, for a deviation $\delta' = 0.05$.
8. Arrangement of a 2D slab magneto-inductive imaging system.
9. a) Dispersion diagram for a loss-less 2D slab magneto-inductive array with $\kappa_x = -0.05$, $\kappa_z = 0.5$ and $N_z = 6$; b) point spread function at input and output, and c) and variation of peak height with position in the slab, for a deviation $\delta'' = 0.001$ above the upper band.
10. Arrangement of a 3D slab magneto-inductive imaging system.
11. Point spread function at a) input and b) output, for a loss-less 3D slab magneto-inductive array with $\kappa_x = \kappa_y = -0.05$, $\kappa_z = 0.5$ and $N_z = 6$, for $\delta''' = 0.001$.
12. Images of the letter 'M' obtained at the output of loss-less 3D arrays with $\kappa_x = \kappa_y = -0.05$, and $\kappa_z = 0.5$ and a) $N_z = 2$, b) $N_z = 4$, and c) $N_z = 6$ for $\delta''' = 0.01$.
13. Point spread function of a lossy 1-D magneto-inductive array, for $\kappa = -0.25$, $\kappa_S = +0.025$, $Q = 1000$ and a) $\delta = 0.05$, b) 0.1 and c) 0.2. In each case, four responses are shown, calculated by ignoring all sources and detectors (Approximate theory), including a single source (Method 1), a single source and a single movable detector (Method 2) and a single source and a line of fixed detectors (Method 3).

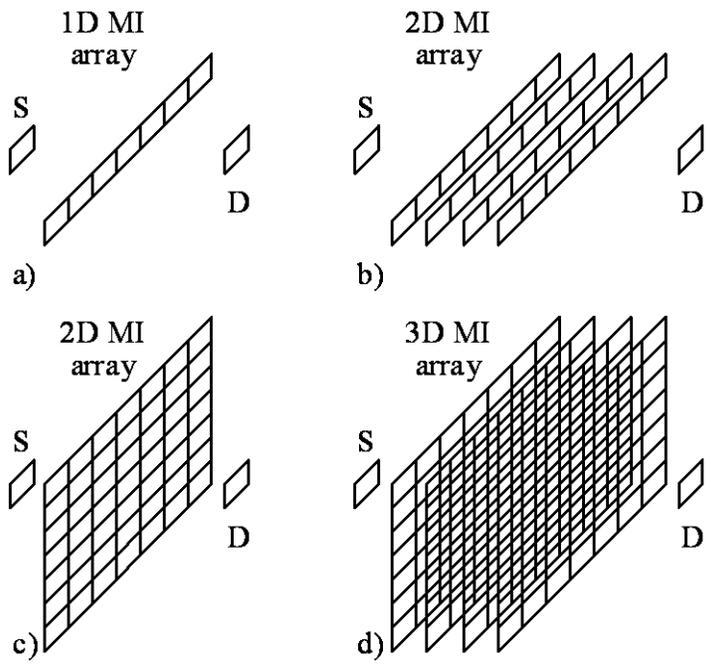
Figure 1.

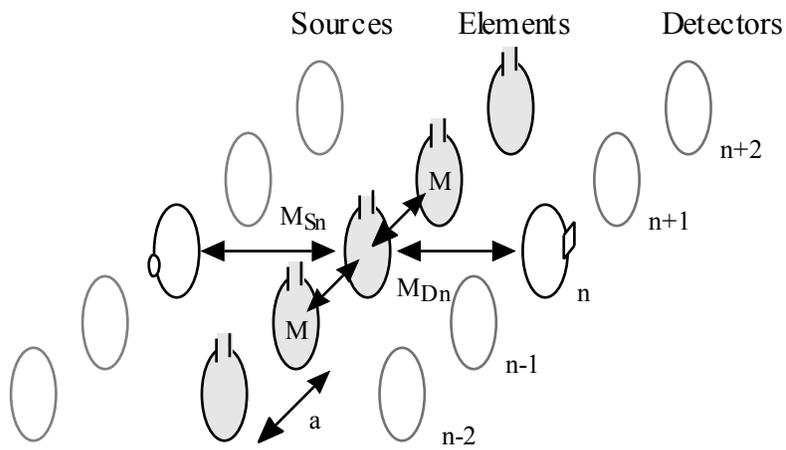

a)

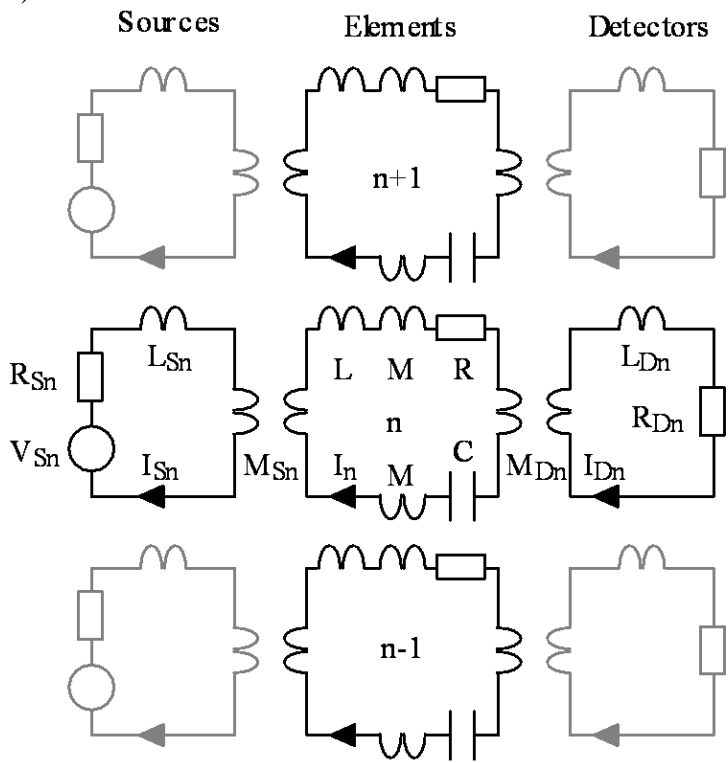

b)
Figure 2.

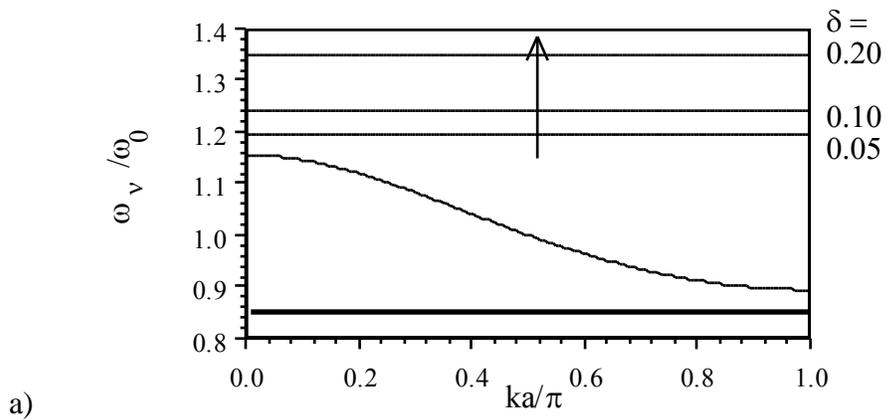

a)

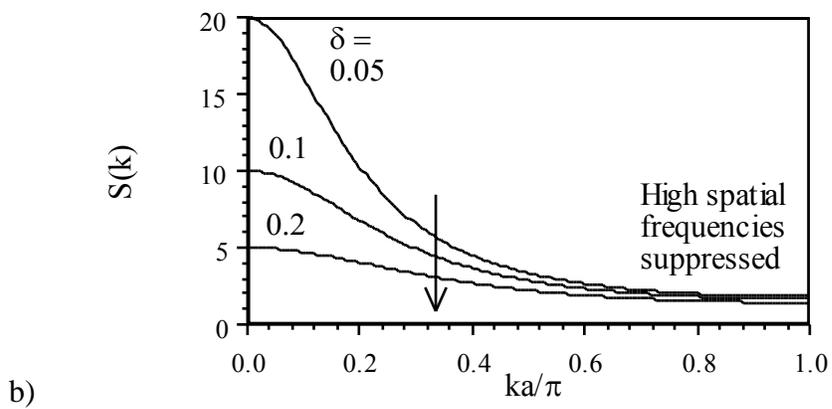

b)

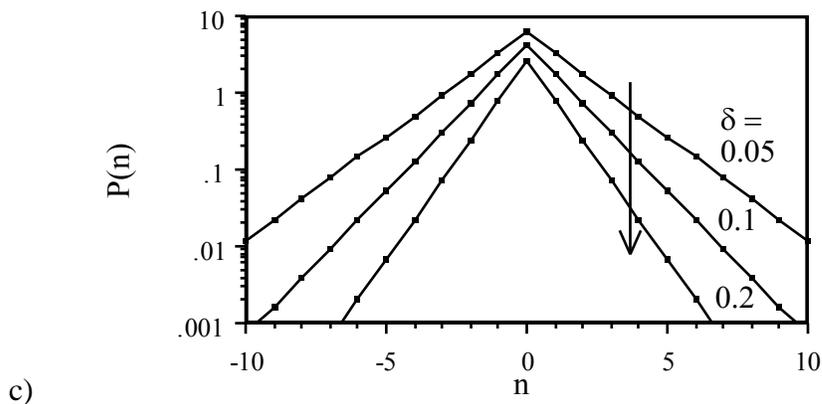

c)

Figure 3.

a)
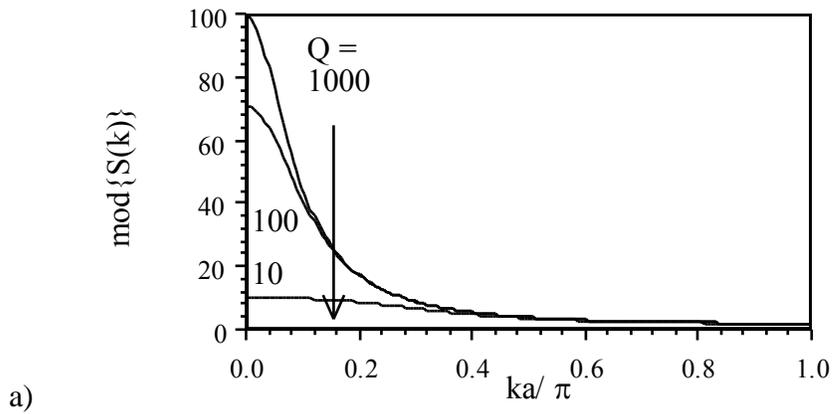

b)
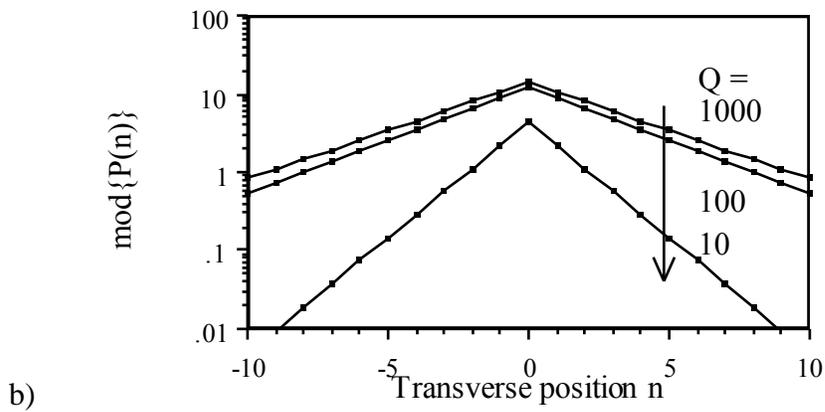

Figure 4.

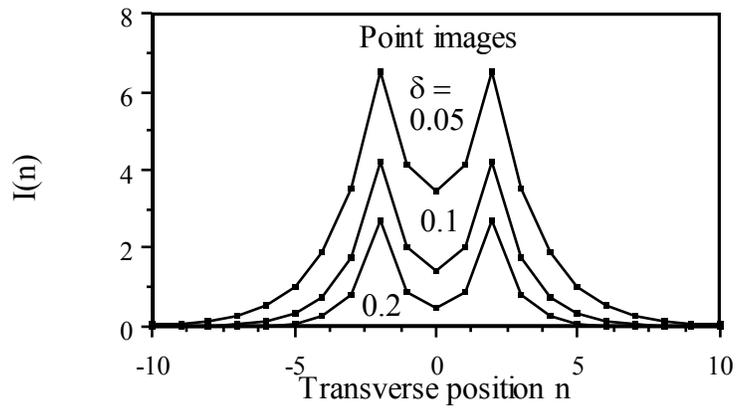

a)

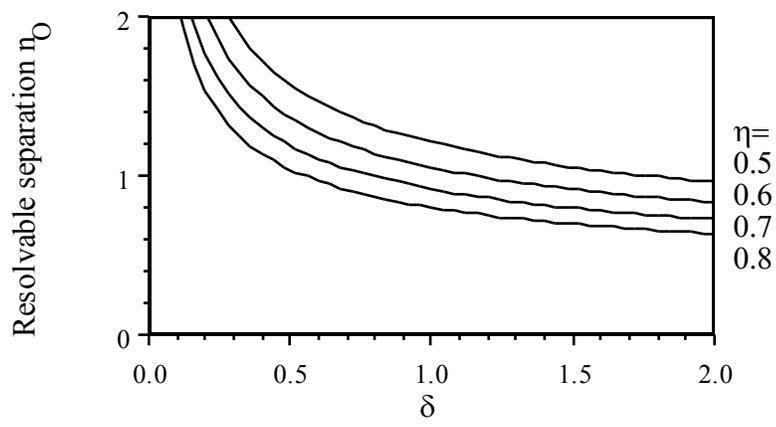

b)
Figure 5.

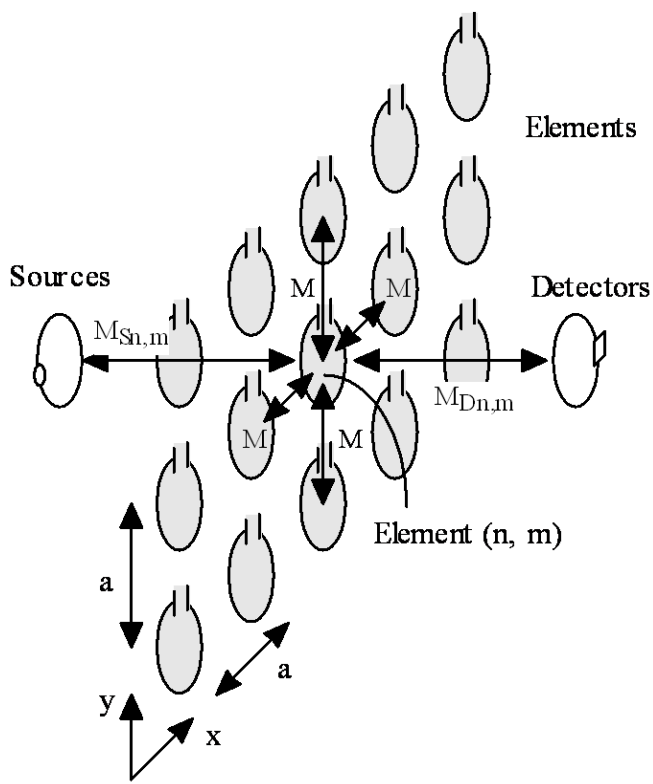

Figure 6.

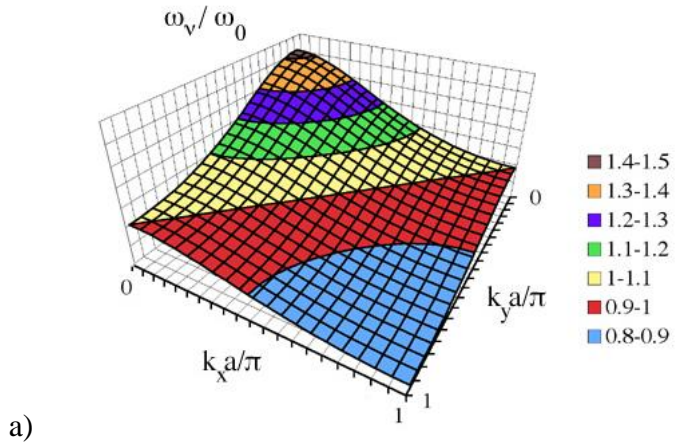

a)

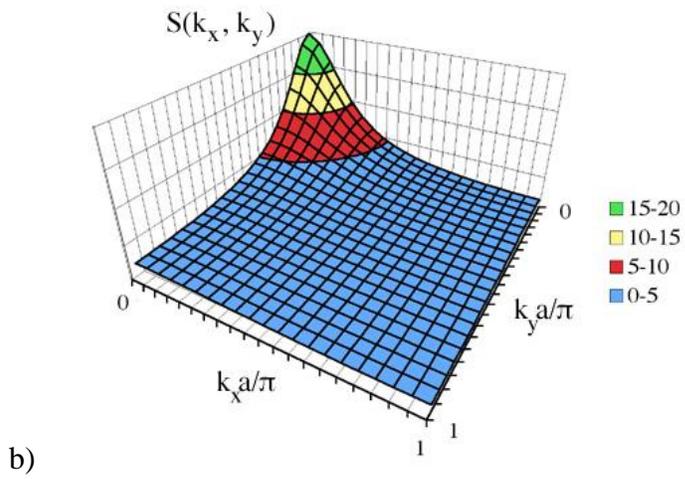

b)

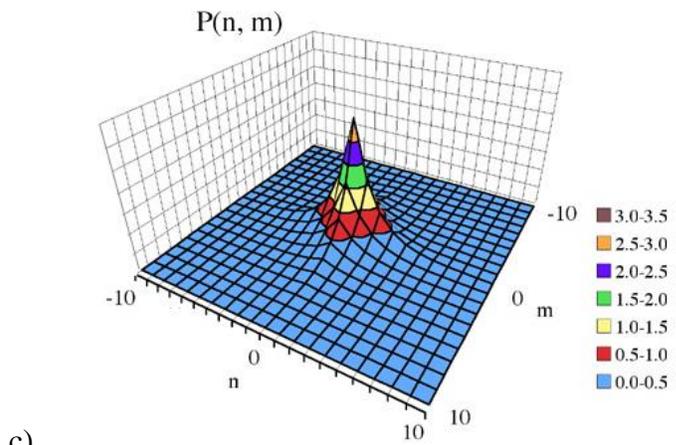

c)
Figure 7.

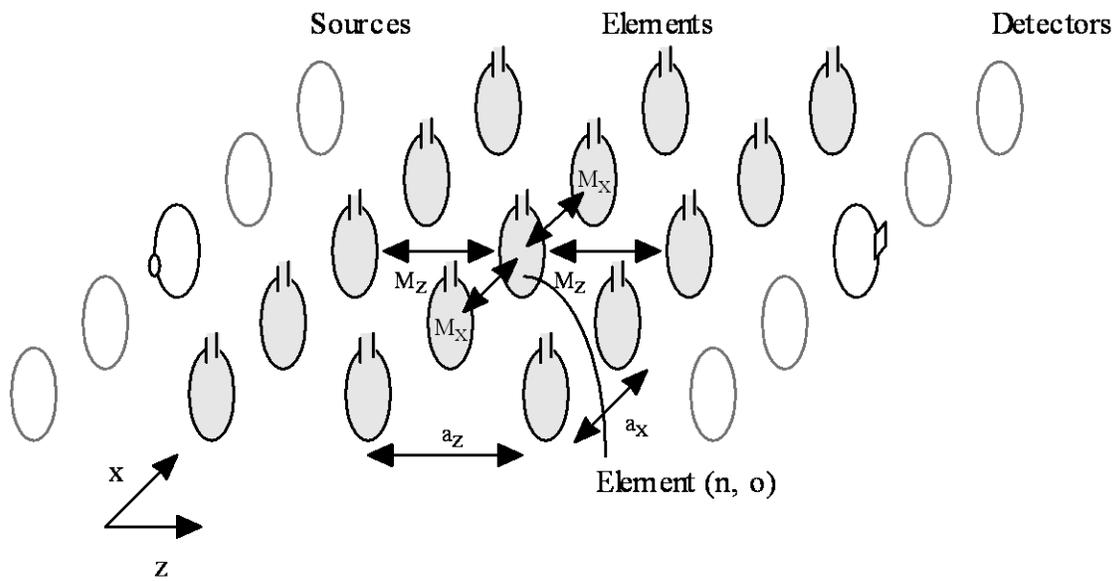

Figure 8.

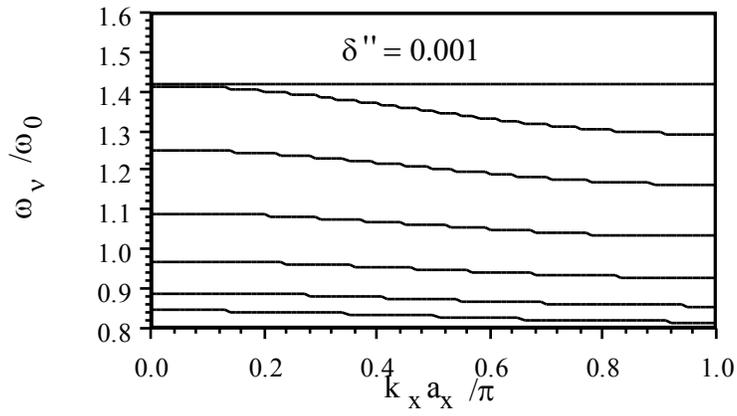

a)

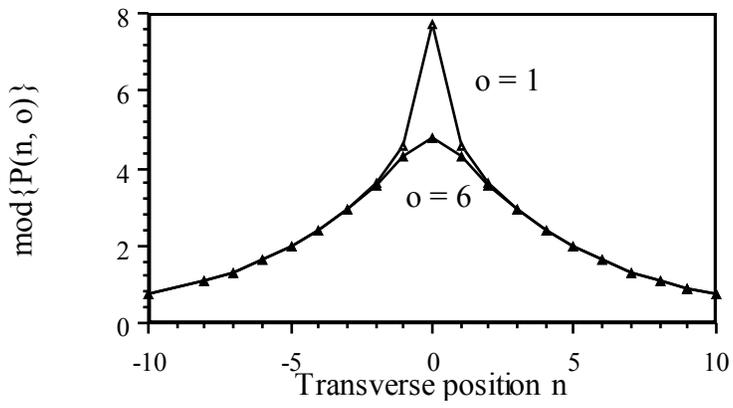

b)

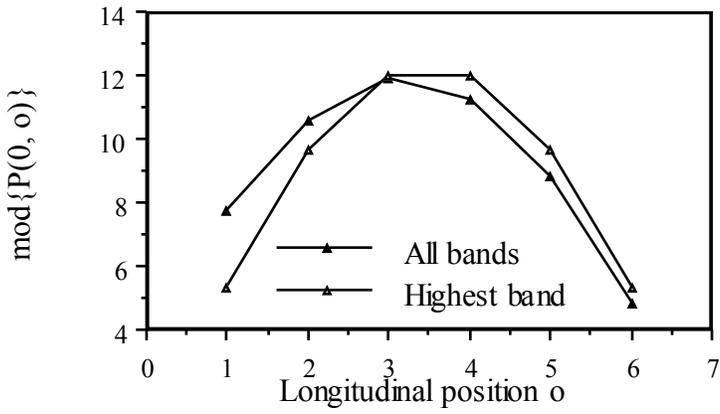

c)

Figure 9.

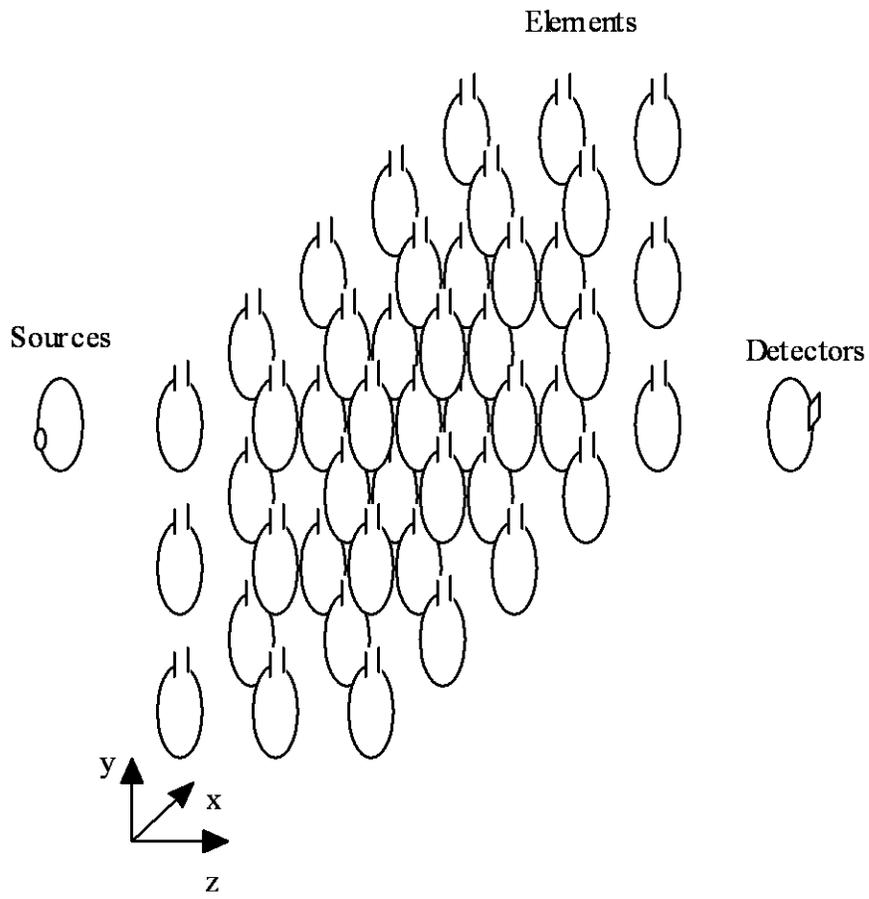

Figure 10.

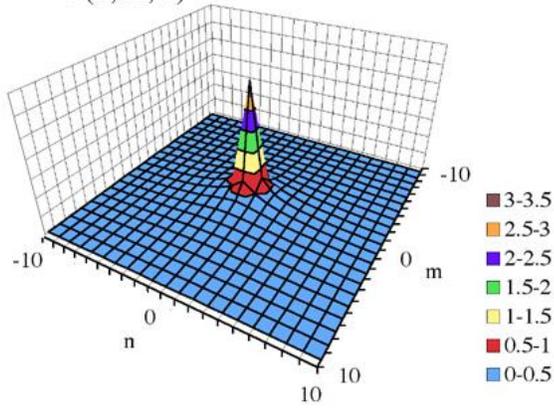

a)

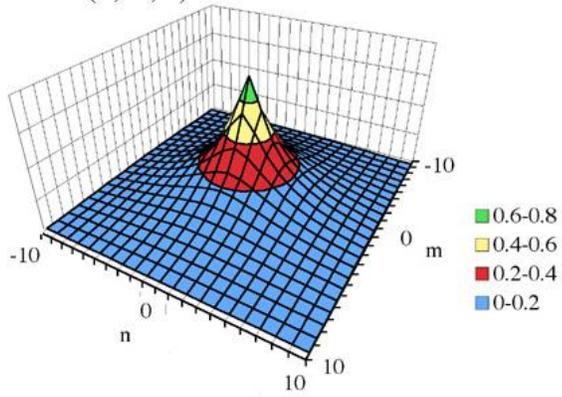

b)

Figure 11.

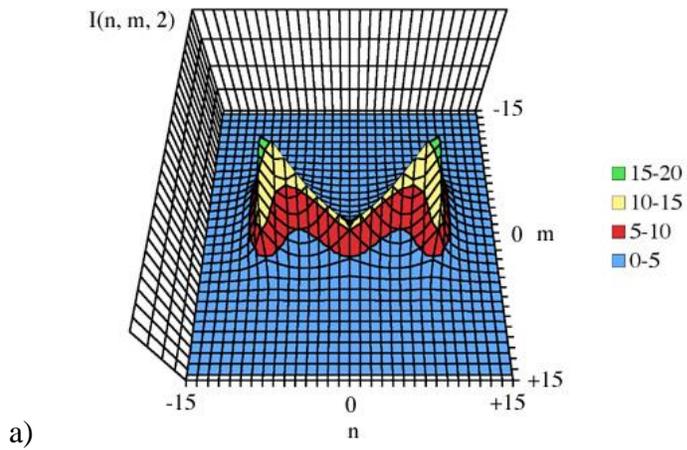
a)

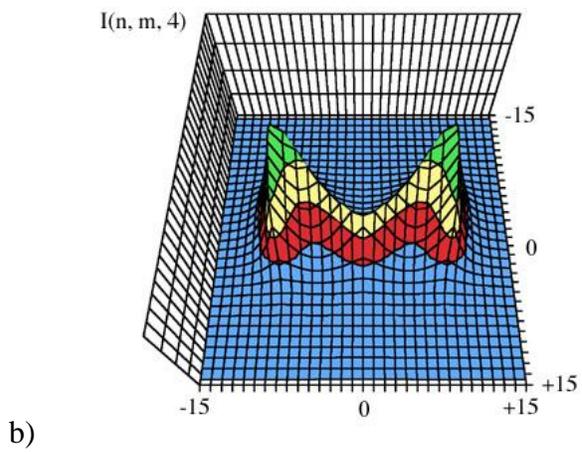
b)

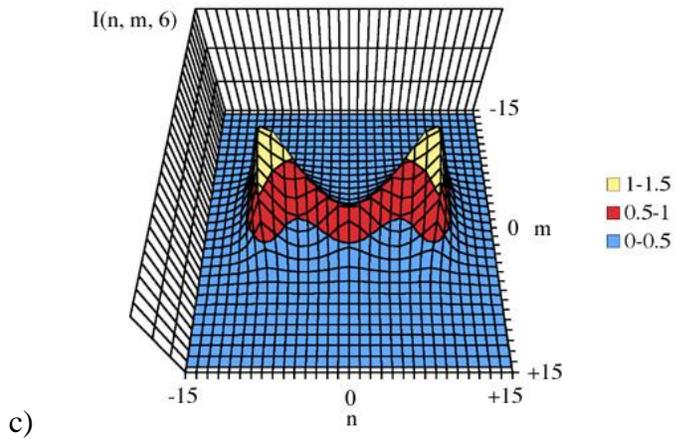
c)
Figure 12.

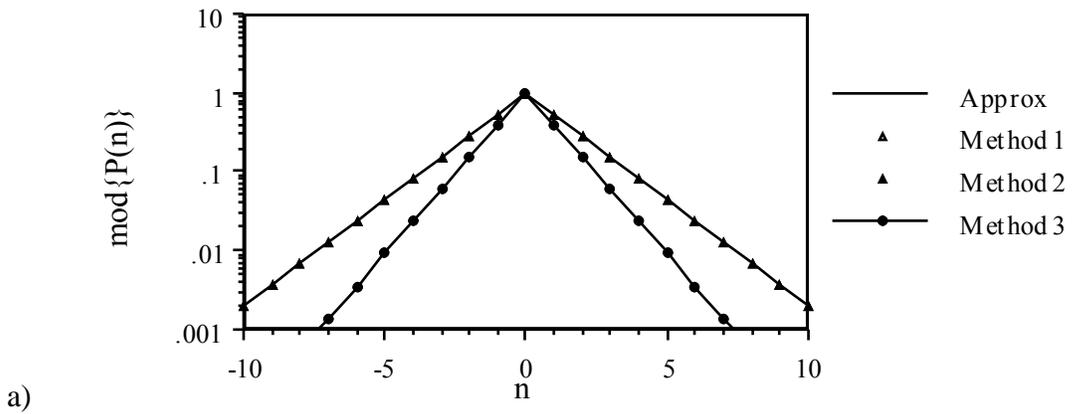

a)

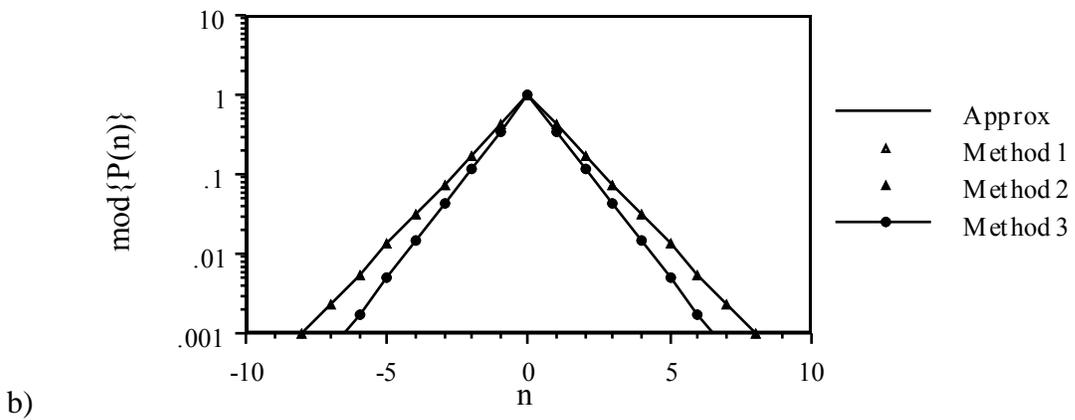

b)

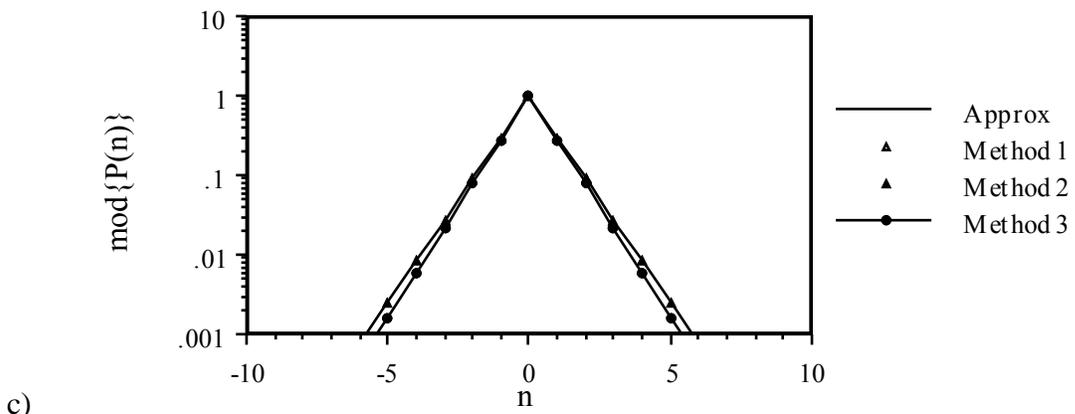

c)

Figure 13.